Degeneracy: a design principle for achieving robustness and evolvability

James Whitacre[a]* and Axel Bender[b]

[a]*Defence and Security Applications Research Centre; University of New South Wales at the Australian Defence Force Academy, Canberra, Australia*

[b] *Land Operations Division, Defence Science and Technology Organisation; Edinburgh, Australia*

*To whom correspondence should be addressed. E-mail: jwhitacre79@yahoo.com  Phone: 61403595280

**Conflicts of Interest**:  The authors have no conflicts of interest to report

**Contributions**:  JW designed and performed research, analyzed data and wrote the paper.  AB analyzed data and wrote the paper.

## *Abstract*

Robustness, the insensitivity of some of a biological system's functionalities to a set of distinct conditions, is intimately linked to fitness. Recent studies suggest that it may also play a vital role in enabling the evolution of species. Increasing robustness, so is proposed, can lead to the emergence of evolvability if evolution proceeds over a neutral network that extends far throughout the fitness landscape. Here, we show that the design principles used to achieve robustness dramatically influence whether robustness leads to evolvability. In simulation experiments, we find that purely redundant systems have remarkably low evolvability while degenerate, i.e. partially redundant, systems tend to be orders of magnitude more evolvable. Surprisingly, the magnitude of observed variation in evolvability can neither be explained by differences in the size nor the topology of the neutral networks. This suggests that degeneracy, a ubiquitous characteristic in biological systems, may be an important enabler of natural evolution. More generally, our study provides valuable new clues about the origin of innovations in complex adaptive systems.

**Keywords**: evolution, fitness landscapes, neutral networks, redundancy, distributed robustness

# 1. Introduction

Life exhibits two unique qualities, both of which are highly desirable and hard to create artificially. The first of these is robustness. At almost every scale of biological organization, we observe systems that are highly versatile and robust to changing conditions. The second quality is the ability to innovate. Here we are referring to life's remarkable capacity for creative, innovative, selectable change. Understanding the origins of robustness, innovation and their relationship, is one of the most interesting open problems in biology and evolution. We begin by briefly describing these concepts and recent progress in understanding their relationship with each other.

## *Robustness – Insensitivity to Varying Conditions*

Despite the numerous definitions of robustness in the literature [1], there is surprising agreement on what robustness means. In its most general form, robustness describes the insensitivity of some functionality or measured system state to a set of distinct conditions. The state is assumed critical to the continued existence of the system, e.g. by being intimately tied to system survival or fitness.

Robustness is a commonly observed property of biological systems [2], and there are many possible explanations for its existence [3] [2] [4]. It is generally agreed that robustness is vital because cells, immune systems, organisms, species, and ecosystems live in changing and often uncertain conditions under which they must maintain satisfactory fitness in order to survive. A biological system can be subjected to both internal and external change. Genotype mutations, variations caused by the stochasticity of internal dynamics, altered species interactions and regime shifts in the physical environment are examples for drivers of such change. Thus, for a population of organisms to be robust, the phenotype needs to be controlled. In some cases, this means maintaining a stable phenotype despite variability of the environment (canalization), while in other cases it requires modification of the phenotype to improve or maintain fitness within a new environment (phenotypic plasticity) [2].

## *Evolvability – Accessibility of Distinct Phenotypes*

Evolvability is concerned with the selection of new phenotypes. It requires an ability to generate distinct phenotypes and a non-negligible selection probability for some of them. Kirschner and Gerhart define evolvability as "an organism's capacity to generate heritable phenotypic variation" [5]. In this sense, evolvability is the dispositional concept of phenotypic variability, i.e. it is the potential or the propensity for the existence of diverse phenotypes [6]. More precisely, it is the total accessibility of distinct phenotypes. As with other studies [7] [8] [9], we use this definition of phenotypic variability as a proxy for a system's evolvability.

Many researchers have recognized the importance of evolvability [5] [8] [9] [10]. By defining natural evolution as 'descent with modification', Darwin implicitly assumed that iterations of variation and selection would result in the successive accumulation of useful variations [10]. However, decades of research involving computer models and simulation have shown that Darwin's principles of natural evolution can only generate adaptive changes that are at best finite and at worst short-lived. It is no longer refuted that the founding principles of evolution are insufficient to evolve systems of unbounded complexity. A modern theory of evolution therefore must unravel the mystery that surrounds the origin of innovations in nature [11] [12].

## *Robustness-Evolvability Paradox*

At first, the robustness of biological systems appears to be in conflict with other demands of natural evolution. On the one hand, species are highly robust to internal and external perturbations while, on the other hand, innovations have evolved continually over the past 3.5 billion years of evolution. Robustly maintaining developed functionalities while at the same time exploring and finding new ones seem to be incompatible.

Progress in our understanding of the simultaneous occurrence of robustness and evolvability in a single system is well illustrated by the recent work of Ciliberti et al [8]. In their study, the authors model gene regulatory networks (GRN). GRN instances are points in the genotypic GRN space and their expression pattern represents an output or phenotype. Together, genotype and phenotype define a fitness landscape. Ciliberti et al discovered that a large number of genotypic changes have no phenotypic effect, thereby indicating robustness to such changes. The phenotypically equivalent systems connect to form a neutral network in the fitness landscape. A search over this neutral network is able to reach genotypes that are almost as different from each other as randomly sampled GRNs. Ciliberti et al found that the number of distinct phenotypes in the local vicinity of the neutral network is extremely large. This indicates that close to a viable phenotype a wide range of different phenotypes can be accessed, leading to a high degree of evolvability. From these results, they propose that the existence of an extensive neutral network can resolve the robustness-evolvability paradox.

The study by Ciliberti et al emphasizes the importance of the neutral network, i.e. the connected graph of equivalent (or nearly equivalent) phenotypes that extends through the fitness landscape. Network connectivity relates directly to robustness because it allows for mutations and/or perturbations that leave the phenotype unchanged [9]. The degree of robustness depends on the local topology and the size of the network. Evolvability, on the other hand, is concerned with long-term movements that can reach over widely different regions of the fitness landscape. An extensive neutral network with a rich phenotypic neighborhood allows evolution to explore many diverse phenotypes without surrendering a system's core functionalities.

Ciliberti et al in [8] were not the first researchers to point to the importance of neutral networks in the evolution of species. Kimura formulated a neutral theory of evolution as early as 1955 [13], which was expanded by Ohta to include nearly neutral conditions [14]. More recently, several other studies have demonstrated the presence of neutral networks in computer models of biological systems. Particularly noteworthy is the pioneering work by Schuster et al [15] who found that neutral networks exist in RNA secondary structures. Ciliberti et al's work, however, is novel because it quantifies phenotypic variability and demonstrates the huge range of accessible phenotypes that can emerge as a consequence of robust phenotypic expression. From this Ciliberti et al conclude (tentatively) that reduced phenotypic variation, i.e. increased mutational robustness, and enhanced phenotypic variability, i.e. increased evolvability, are positively correlated in natural evolution. The topology of the neutral network, so they suggest, may matter greatly.

## *Redundancy and Degeneracy – Design Principles for Robustness*

Redundancy and distributed robustness are two basic design principles that are believed to play an important role in achieving robustness in biological systems [16] [17]. Redundancy is an easily recognizable design principle that is prevalent in both biological and man-made systems. Here, redundancy means 'redundancy of parts' and refers to the coexistence of identical components with identical functionality. It is a common feature of engineered systems where redundancy provides robustness against variations of a very specific type ('more of the same' variations). For example, redundant parts can substitute others that malfunction or fail, or augment output when demand for a particular output increases. Redundancy is also prevalent in biology. Polyploidy, as commonly found in fern, flowering plant or some lower-form animal eukaryotic cells, homogenous tissues and allozymes are examples of functional biological redundancy. Another and particular impressive example is neural redundancy, i.e. the multiplicity of neural units (e.g. pacemaker cells) that perform identical functions (e.g. generate the swimming rhythms in jellyfish or the heartbeat in humans). For instance, almost half of the parvocellular axons in the human optic nerve appear to be redundant.

Distributed robustness emerges through the actions of multiple dissimilar parts [17] [18]. It is in many ways unexpected because it is only derived in complex systems where heterogeneous components (e.g. gene products) have multiple interactions with each other. In our experiments we show that distributed robustness can be achieved through degeneracy (see Section 2). Degeneracy is also known as partial redundancy. In biology it refers to conditions under which the functions or capabilities of components overlap partially [16]. It particularly describes the coexistence of structurally distinct components (genes, proteins, modules or pathways) that can perform similar roles or are interchangeable under certain conditions, yet have distinct roles under other conditions.

Degeneracy is ubiquitous in biology as evidenced by the numerous examples provided by Edelman and Gally [16]. One case in point is the adhesins gene family in *Saccharomyces*, which expresses proteins that typically play unique roles during development, yet can perform each other's functions when their expression levels are elevated [19]. Another example is found in glucose metabolism which can take place through two distinct pathways, glycolysis and the pentose phosphate pathway, that can substitute for each other if necessary [20]. Ma and Zeng argue that the robustness of the bow-tie architecture they discovered in metabolism is largely derived through the presence of multiple distinct routes to achieving a given function or activity [21].

## *Fitness landscape model*

In this study, we investigate how redundant and degenerate design principles influence a system's robustness and evolvability, and whether any observable differences can be accounted for by the properties of the neutral network. We use an exploratory abstract model of a fitness landscape that is designed with the following considerations in mind. First, the model should enable an unambiguous distinction between redundant and degenerate systems. Second, interactions between components should be simple in order to explore the mechanistic differences between redundant and degenerate robustness. Third, we want a model that is minimalist in the conditions needed to probe our research questions. By pursuing this minimalist approach, we sacrifice some biological fidelity. For instance we constrain our study to a linear genome-proteome model; however, we believe that the general principles we explore apply broadly to other biological systems and abiotic complex adaptive systems. Indeed, our aim is to arrive at conclusions that are widely applicable to systems that are subjected to variation and selective driving forces and thus need to be both robust and evolvable in order to maintain long-term viability.

**Model Overview**
We model the genotype-phenotype map of a genetic system neglecting population properties, i.e. we explore evolution over a fitness landscape with a population size of one. In our model, each gene expresses a single protein product that has multiple functional targets (e.g. non-trivial interactions with multiple molecular species). Through these interactions a gene product contributes to the formation of multiple phenotypic traits (pleiotropy) In the mapping from genotype to phenotype in biology, a gene's influence on different traits can vary in the size of its effect and these effects can either be related or functionally separated, e.g. through spatially and temporally isolated events. In our model, we simplify this aspect of the mapping and assume that each functional target of a gene product influences a distinct (separable) trait and furthermore that each of these events is additive and has the same effect size (see Figure 1). This abstraction in modelling gene pleiotropy is partly justified because of 1) the knowledge that many proteins act as versatile building blocks and can perform different cellular functions with the function depending on the complex a protein forms with other gene products [22] [23]; and 2) evidence that the scope of protein multi-functionality is broad [24] [25] and the execution of these functions typically occurs at different times [26].

The gene products we model are energetically driven to form complexes with a genetically predetermined set of functional target types. However, through the mediated availability of targets, each functionally versatile gene product can vary in how it contributes to system traits. This variation will depend on the genetic and environmental background, i.e. what targets are available to bind with and what other gene products it must compete with due to the functional overlap of gene products. This competition among gene products enables compensatory actions to occur within the model.

**Phenotype Attractor**

We assume that the architecture of any genome-proteome mapping is such that it spontaneously organizes towards a set of stable trait values (homeostasis), and that the attractor for these system dynamics is robust either as a consequence of its evolutionary history (genetically canalized) or due to system properties that generically lead to capacitance and buffering, e.g. see [27] [3]. In a deliberate departure from other studies, we do not explicitly simulate regulatory interactions that direct system dynamics towards this phenotypic attractor, e.g. where protein-target binding events directly regulate the composition of proteins and targets expressed in the system. Instead, we assume that an ancestral phenotype represents a strong 'stationary' attractor for the system. This does not preclude the phenotypic traits themselves from being non-stationary. The separation between the phenotypic attractor and the components that comprise the system, although rarely considered in simulations, allows for interesting insights into the material conditions that limit phenotypic control.

Thus when this system finds itself in a perturbed phenotypic state (i.e. new targets or new proteins), compensatory actions by extant gene products are taken, if available, that move the phenotype towards its attractor. In our model, these actions simply consist of changes in protein-target binding that are made based on the availability of functional targets and competition between functionally redundant gene products. One consequence of this model is that adaptive genetic mutations are only possible when mutations prevent the system from accessing its ancestral attractor. As we will demonstrate, the exposure of new phenotypes is ultimately influenced by what previously appears as cryptic genetic changes. Below we give a concise description of the parameters and functions defining a mathematical realization of this model.

**Technical Description:** The model consists of a set of genetically specified proteins (i.e. material components). Protein state values indicate the functional targets they have interacted with and also defines the trait values of the system. The genotype determines which traits a protein is able to influence, while a protein's state dictates how much a protein has actually contributed to each of the traits it is capable of influencing. The extent to which a protein $i$ contributes to a trait $j$ is indicated by the matrix elements $C_{ij} \in Z$. Each protein has its own unique set of genes, which are given by a set of binary values $\delta_{ij}$, $i \in n$, $j \in m$. The matrix element $\delta_{ij}$ takes a value of one if protein $i$ can functionally contribute to trait $j$ (i.e. bind to protein target $j$) and zero otherwise. In our experiments, each gene expresses a single protein (no alternative splicing). To simulate the limits of functional plasticity, each protein is restricted to contribute to at most two traits, i.e. $\sum_{i \in n} \delta_{ij} \leq 2 \ \forall i$. To model limits on protein utilization (i.e. caused by the material basis of gene products), maximum trait contributions are defined for each protein, which for simplicity are set equal, i.e. $\sum_{j \in m} C_{ij} \delta_{ij} = \lambda \ \forall i$ with the integer $\lambda$ being a model parameter.

The set of system traits defines the system phenotype with each trait calculated as a sum of the individual protein contributions $T_j^P = \sum_{i \in n} C_{ij} \delta_{ij}$. The environment is defined by the vector $T^E$, whose components stipulate the number of targets that are available. The phenotypic attractor $F$ is defined in Eq. 1 and acts to (energetically) penalize a system configuration when any targets are left in an unbound state, i.e. $T_j^P$ values fall below the satisfactory level $T_j^E$. Through control over its phenotype a system is driven to satisfy the environmental conditions. This involves control over protein utilization, i.e. the settings of $C$. We implement *ordered asynchronous updating* of $C$ where each protein stochastically samples local changes in its utilization (changes in state values $C_{ij}$ that alter the protein's contribution to system traits). Changes are kept if compatible with the global attractor for the phenotype defined by Eq. 1. Genetic mutations involve modifying the gene matrix $\delta$. For mutations that cause loss of gene function, we set $\delta_{ij} = 0 \ \forall j$ when gene $i$ is mutated.

$$F(T^P) = -\sum_{j \in m} \theta_j$$

$$\theta_j = \begin{cases} 0, & T_j^P > T_j^E \\ (T_j^P - T_j^E), & else \end{cases}$$

(1)

We model degeneracy and redundancy by constraining the settings of the matrix $\delta$. This controls how the trait contributions of proteins are able to overlap. In the '*redundant model*', proteins are placed into subsets in which all proteins are genetically identical and thus influence the same set of traits. However, redundant proteins are free to take on distinct state values, which reflects the fact that proteins can take on different functional roles depending on their local context. In the '*degenerate model*', proteins can only have a partial overlap in what traits they are able to affect. The intersection of trait sets influenced by two degenerate proteins is non-empty and truly different to their union. An illustration of the redundant and degenerate models is given in Figure 1.

## *Measuring Robustness and Evolvability*

Using the model just described, we investigate how the functional overlap in genes places theoretical limits on a system's capacity to regulate its phenotype. This involves evaluating the latent canalization potential of a system (i.e. measuring robustness) as well as the uniqueness of phenotypes associated with evolutionarily accessible genotypes that are not fully canalized (i.e. measuring evolvability). Many of the analytical concepts and steps are similar to those in the study of gene regulatory networks by Ciliberti et al [8] and RNA secondary structure in [11].

**Fitness landscape.** First, we consider a network in which each node represents a particular system–environment tuple. The tuple is characterized by $\delta$ (genotype), $C$ (phenotype) and $T^E$ (environment). Connections (arcs) represent feasible variations in conditions, i.e. external or internal changes to respectively $T^E$ or $\delta$. Each node can be assigned a fitness value according to Eq. 1; thus the network is a generalized representation of a fitness landscape. For simplicity, we assume that all changes that occur along arcs are equally probable and reversible, e.g. we neglect variability in mutation rates across the genome. The result of this assumption is an unweighted and undirected network, for which the robustness and evolvability calculations are simplified. In response to a genetic mutation, i.e. a one-step move within the network, the phenotype is subjected to ordered asynchronous updating that is driven by the phenotypic attractor of the system (Eq. 1).

**Neutral network.** A neutral network is defined as a connected graph of nodes with equal fitness. One can consider it a connected set of external and internal conditions within which a system has the same fitness. Notice that connectedness implies that each node within the network can be reached by every other without changing the system's fitness along the path of arcs. Ignoring population properties, this implies selective neutrality. We restrict the class of condition changes to single gene mutations, allowing us to recover Ciliberti et al's neutral networks [8], which exist within a fitness landscape such as originally described by Wright [28]. Compared with [8], we relax the neutrality criterion slightly. We consider all systems neutral that are within $\alpha$ % of the original system fitness. This relaxation is necessary to describe "satisficing behavior". Justifications for approximate neutrality are varied in the literature. Typically, they are based upon constraints that are observed in physical environments and lead to reductions in selection pressure or limitations to perfect selection.

**1-neighborhood and evolvability.** Similar to [9], we define a 1-neighborhood of all non-neutral nodes that are directly connected to a neutral network. These nodes represent mutations that, for the first time, result in non-neutral changes of system fitness. As defined in Section 1 evolvability is equivalent to the total accessibility of distinct phenotypes. It thus equals the count of unique phenotypes in the 1-neighborhood.

**Robustness** can be evaluated in many ways, and we therefore introduce several robustness metrics. For a system (i.e. a node) in the neutral network, its *local robustness* is defined as the proportion of arcs that connect it to the neutral network. In other words, local robustness is the proportion of immediately possible (single gene) mutations under which a system can maintain its fitness. The local robustness measurements reported in the next section are the local robustness for each neutral node averaged over all neutral nodes. An alternative robustness measure is a system's *versatility*. It is the total count of distinct and mutationally accessible genotypes under which a system can remain sufficiently fit. Since we assume that the network of changes is unweighted and undirected, versatility is directly proportional to, and thus well approximated by, the size of the neutral network. Finally, we measure *differential robustness* by analyzing a system's response to increasingly larger mutation rates. For this we record the fitness of the initial genetic system as it is subjected to increasingly large numbers of genetic mutations.

**Fitness landscape exploration.** In order to measure evolvability, both the neutral network and the 1-neighborhood need to be explored. The details of the algorithm for searching the fitness landscape is given in Section 4. Unless stated otherwise, the remaining experimental conditions are observed in all experiments. Genotypes $\delta_{ij}$ are randomly initialized as binary values that meet the previous section's constraints, including the requirements of degeneracy or redundancy as of the model being tested. For the initial system (first node in the neutral network), component state values $C_{ij}$ are randomly initialized as integer values between 0 and $\lambda$. The initial environment $T^E$ is defined as the initial system phenotype. The neutrality threshold is set to $\alpha = 5\%$ and the model parameter $\lambda$ to $\lambda = 10$. The number of traits is $m = 8$, and the number of system components $n = 2m = 16$. In our random initializations of the models we enforce that each trait has the exact same number of proteins contributing to it; thus $\sum_{j \epsilon m} \delta_{ij} \equiv 4$. This ensures that the redundant and degenerate models start with systems that have exactly the same fitness and functionalities. Furthermore, the size of the fitness landscape and gene mutations have been defined to be identical for both types of models. Ad hoc experiments varying the settings of $\lambda$, $n$, and $m$ did not alter our basic findings. Each experiment is conducted with 50 experimental replicates.

Based on several considerations, we decided not to explicitly model genetic mutations that create novel functions or model the recruitment of gene products to previously unrelated system traits, i.e. changes to protein specificity. First, this would require us to make additional assumptions about the topology of protein functional space and the selective relevance of new functions within an environment. Secondly, for almost any protein function landscape one could envision, increasing the number of points sampled in the landscape increases the mutational accessibility of distinct functions, up to saturation. Because the degenerate model displays greater gene diversity (a requirement derived from the definition of redundancy) we wanted to remove any confounding effects that could be caused by differences in the number of distinct genes in the two systems. If we had allowed for genetic mutations other than loss of function, the observed differences in system evolvability that are presented in our results would have been off-handedly attributed to differences in mutational access between the two fitness landscapes.

## 2. Results

### *Design principles considerably affect system evolvability*

First, we investigate how the system design principles influence robustness and evolvability. In Figure 2 we show results for local robustness, versatility and evolvability as the algorithm explores the neutral networks and 1-neighborhoods. Presenting the results in this way exposes the rate at which new neutral genotypes and new (non-neutral) phenotypes are being discovered during the search process. Over the evolution of the networks, a degenerate system is found to be over twice as versatile as a redundant system, with the neutral network sizes converging to respectively $NN_{deg} = 660 + 15 - 50$ and $NN_{red} = 280 \pm 5$ after $3 \times 10^5$ search steps. This means that a degenerate system maintains sufficient fitness in approximately twice as many circumstances of gene deletions as a redundant system. After $3 \times 10^5$ search steps the 1-neighbourhood of the degenerate system contains $1{,}900 + 600 - 400$ unique phenotypes compared with merely $90 \pm 30$ for the redundant system. Thus, the degenerate system is about 20 times more evolvable than the redundant system. The local robustness of the two systems is initially quite different ($R_{deg} = 0.38 \pm 0.005$, $R_{red}= 0.30 \pm 0.005$) with the difference becoming smaller (but remaining significant, $p < 1E-6$) as the exploration of the neutral networks progresses. This indicates that the robustness of the initial degenerate system is much larger, however the average robustness of all genotypes on the neutral network is less distinct. In other words, the robustness advantage for the degenerate system is reduced as the neutral network is being explored. A similar effect is expected in natural evolution as populations are driven towards mutation-selection balance and subsequently towards less robust regions of the network.

To obtain a clearer sense of the robustness of each design principle, we analyze the differential robustness of the initial (un-mutated) systems when being subjected to increasingly larger numbers of mutations. Differential robustness is given as the proportion of conditions for which the perturbed systems can maintain satisfactory fitness. Unlike in the experiments shown in Figure 2 we do not only explore the effect of single gene mutations but also that of multiple gene mutations. Figure 3 demonstrates that, on average, degenerate systems are more robust to increasingly larger changes in conditions than are redundant systems.

Our experiments strongly support the finding that the design principles markedly influence system properties such as neutral network size, robustness and evolvability. It is not clear however, why the evolvability of these systems is so dramatically different. In particular, it is not clear whether the differences in evolvability can be accounted for by differences in robustness or in neutral network size. In the remaining experiments, we explore these questions and show that differences in neutral network size, topology, and system robustness cannot account for the huge differences in evolvability. From this, we are left to conclude that the design principles are mainly responsible for the observed effect.

### *Evolvability does not derive from neutral network exploration*

In the next set of experiments, we evaluate the properties of the neutral networks (that are encountered during the search process) to determine if these can account for the observed differences in system evolvability. First we check if the size of the explored neutral network is the main determinant of the number of unique phenotypes that are discovered. In Figure 4 we see that large neutral networks do not necessarily lead to a greater access to unique phenotypes. The redundant systems are strongly limited with respect to their accessibility of distinct phenotypes. We only observe a small dependence of evolvability on the size of the neutral network. In the degenerate system, on the other hand, the exact opposite is observed: accessibility of new unique phenotypes increases considerably as new regions of the neutral network are explored.

### *Neutral network topology cannot account for evolvability*

If the size of the explored neutral network is not highly correlated with the observed differences in evolvability, it seems reasonable to suspect that the manner in which the neutral network extends across genotype space could influence system evolvability, as suggested in [8]. To test this hypothesis, we analyze network distances, i.e. proxies for a network's ability to reach distinct regions of genotype space.

One way of determining the distance between two nodes is to calculate the shortest path, or "geodesic", between them. If we take, for a specific node, the average of the geodesics to all other nodes in the network, and then take the average of these over all nodes in the network, we get the so called *characteristic path length*. In panel a) of Figure 5 we show this characteristic path length as a function of network size. Characteristic path length increases with network size and approaches largely similar values at NN=800 for the two design principles (12.4 for degenerate and 10.8 for redundant systems). The small differences in characteristic path length though do not explain the huge differences in evolvability as shown in panel b) of Figure 5. Similar conclusions can be drawn when other network distance measures are analyzed, such as the top 10% longest path lengths or the Hamming distance in genotype space (see Figure 6).

### *Versatility and local robustness do not guarantee evolvability*

In the next set of experiments, we investigate whether versatility and local robustness can account for differences in evolvability. For this, we study the effect of making available additional resources while maintaining environmental (trait) requirements, i.e. maintaining the same phenotypic attractor. We employ the same experimental conditions as previously, with the exception that we increase the number $n$ of system genes that can be expressed and that can contribute to system traits. Due to their additive effect on system traits, the inclusion of new functional genes should make both types of systems – redundant and degenerate – more robust to loss of function gene mutations and act to establish larger neutral networks.

As shown in Figure 7, adding excess functional genes indeed increases the size of the neutral network as well as the local robustness for both types of systems. Surprisingly however, the redundant system does not display a substantial growth in evolvability. The degenerate system, on the other hand, is found to have large increases in evolvability and becomes orders of magnitude more evolvable than the redundant model, even when $n$ increases only modestly. The most important conclusion we draw from this is that neither local robustness nor versatility can guarantee that a system will be highly evolvable. This fact can be directly observed in Figure 8 where, for the two system designs, the evolvability data of Figure 7 are plotted as functions of versatility and local robustness.

## 3. Discussion

Taken as a whole, our results indicate that the mechanisms used to achieve robustness generally determine how evolvable a system is. In particular, we showed that differences in the evolvability of a fitness landscape are not necessarily due to differences in local robustness, versatility (neutral network size) or neutral network topology. Mutational robustness and neutrality achieved through redundancy alone does not lead to evolvable systems, regardless of the size of the neutral network. On the other hand, robustness that is achieved through degeneracy can dramatically increase the accessibility of distinct phenotypes and hence the evolvability of a system. Using evidence from biological studies, Edelman and Gally were the first to propose that degeneracy may act both as a source of robustness and innovation in biological systems [16]. Here we have provided the first experimental evidence that supports this relationship between degeneracy and evolvability. However, from observing how evolvability scales with system size in the two classes of models considered in this study, we conclude that degeneracy does not only contribute to new innovations, but that it could be a precondition of evolvability.

## *Degenerate distributed robustness*

How degeneracy allows for distributed robustness and evolvability is not obvious however our model was designed in order to help explore this issue. Below we illustrate how degeneracy creates a connectivity of buffering actions in our model where stress that originates from localized perturbations is diffused to other parts of the system, even though the actions and the functional properties of those actions are not all identical. Such diffusion through networked buffering could be a new source of distributed robustness in biological systems. We emphasize that this buffering network would not have been easy to observe had the phenotypic attractor been an endogenous (self-organized) property of the system.

An illustration of our hypothesis is given in Figure 9. In this illustration, clusters of nodes represent functional groups that contribute to the phenotypic traits of the system. If a particular functional group is stressed (e.g. through loss of contributing components or changes in the desired trait value), the degeneracy of components allows resources currently assigned to other functional groups to be reassigned and alleviate this stress ("buffering"). Depending on the topology of buffer connectivity and the current placement of resources, excess resources that were initially localized can quickly spread and diffuse to other regions of the system. Small amounts of excess functional resources are thus found to be much more versatile; although interoperability of components is localized, at the system level resources can be seen to have huge reconfiguration options. The buffer connectivity and associated reconfiguration options are clearly not afforded to the redundant system (see Figure 9b).

In the degeneracy models considered in our study, the number of links between nodes is constant but otherwise randomly assigned. This random assignment results in a small-world effect in the buffering topology, such that careful design is not necessary to ensure the connectivity of buffers. Hence, the distributed robustness effect is expected to be germane to this class of systems. Future studies will investigate whether constraints imposed by the functional landscape of components can limit the level of distributed robustness observed from degeneracy.

## *The role of degeneracy in evolution*

The investigation of our abstract model provides new clues about the relationship between degeneracy, robustness, and evolution. In the gene deletion studies presented here, we found that the degenerate system can reach a desired phenotype from a broad range of distinct internal (genetic) conditions. We have conducted parallel experiments involving changes to the environment (incremental changes to $T^E$) that have found that the degenerate system can also express a broad range of distinct phenotypes from the same genotypic makeup. Taken together, these results outline two complementary reasons for why distributed robustness can be achieved in degenerate systems. In particular, we speculate that it is both the diversity of unique outputs (i.e the potential for phenotypic plasticity) in addition to the multitude of ways in which a particular output can be achieved (i.e the potential for canalization) that allows for distributed robustness in degenerate systems. Although this richness in phenotypic expression increases the number of unique ways in which the system can fail, it also opens up new opportunities for innovation. Hence, degeneracy may afford the requisite variety of actions that is necessary for both robustness *and* system evolvability. These plasticity and canalization properties in the genotype-phenotype mapping are unique to the degenerate system and moreover are generally consistent with studies of cryptic genetic variation in natural populations.

The evolution of complex phenotypes requires a long series of adaptive changes to take place. At each step, these adaptations must result in a viable and robust system but also must not inhibit the ability to find subsequent adaptations. Complexity clearly demands evolvability to form such systems and robustness to maintain such systems at every step along the way. How biological systems are able to achieve these relationships between robustness, evolvability and complexity is not known. However it is clear that the mechanisms that provide robustness in biological systems must at the very least be compatible with occasional increases in system complexity and must also allow for future innovations. We believe that degeneracy is a good candidate for enabling these relationships in natural evolution. As already noted in this study, degeneracy is unique in its ability to provide high levels of robustness while also allowing for future evolvability. Moreover, in [29] it was found that only systems with high degeneracy are also able to achieve high levels of hierarchical complexity, i.e. the degree to which a system is both functionally integrated and locally segregated [29]. Based on these findings and other supporting evidence illustrated in Figure 10 and summarized in Table 1, we speculate that degenerate forms of robustness could be unique in their capacity to allow for the evolution of complex forms.

It has been proposed that the existence and preservation of degeneracy within distributed genetic systems can be explained by the Duplication-Degeneracy-Complementation model first proposed in [30]. In the DDC model, degenerate genes are retained through a process of sub-functionalization, i.e. where a multi-functional ancestral gene is duplicated and these duplicate genes acquire complementary loss-of-function mutations. Although the present study does not consider the origins of genetic degeneracy or multi-functionality, our results do suggest alternate ways by which degeneracy could be retained during evolution. First, we have shown that degeneracy amongst multifunctional genes has a positive and systemic effect on robustness that is considerably stronger than what is achieved through pure redundancy. Under conditions where the acquisition of such robustness is selectively relevant, e.g. due to variable conditions inside and outside an organism, degenerate genes could be retained due to a direct selective advantage. In this scenario, the ubiquity of degeneracy would be due to its efficacy as a mechanism for achieving selectively relevant robustness, while its impact on evolvability and its compatibility with hierarchical complexity could lead to the emergence of increasingly complex phenotypes.

Alternatively, the enhanced robustness from degeneracy may facilitate its preservation even without a direct selective advantage. Newly added degenerate genes can increase the total number of loss of function mutations with no phenotypic effect, however without a selective advantage this enhanced robustness can be subsequently lost under mutation-selection balance. Due to the distributed nature of the robustness provided, neutral mutations will emerge in several genes that are functionally distinct from the degenerate gene. Following one of these mutations, the compensatory effects of the degenerate gene would be revealed, making its continued functioning selectively relevant and thereby ensuring its future retention within the genetic system. Considering the large mutational target represented by these other genes, retention of the degenerate gene would be a likely outcome in this scenario. Thus, degeneracy may become ingrained within genotypes precisely due to the distributed nature of their compensatory effects, even if these effects do not initially have a selective relevance.

## 4. Methods

*Neutral Network Generation*
Starting with an initial system and a given external environment, defined as the first node in the neutral network, the neutral network and 1-neighborhood are explored by iterating the following steps: 1) select a node from the neutral network at random; 2) change the conditions (genetic mutation or change in environment) based on the set of feasible transitions; 3) allow the system to modify its phenotype in order to robustly respond to the new conditions; and 4) if fitness is within $\alpha$ % of initial system fitness then the system is added to the neutral network, else it is added to the 1-neighborhood of the neutral network.

Additions to the neutral network and 1-neighborhood must represent unique conditions, i.e. $(T^E,\delta)$ pairs, meaning that duplicate conditions are discarded when encountered by the search process. The sizes of the neutral network and 1-neighborhood can be prohibitively large to allow for an exhaustive search and so the neutral network search algorithm includes a stopping criterion after $3 \times 10^5$ steps (changes in condition).

*Results- neutral shadow*
The neutral shadow results are obtained by running the neutral network and 1-neighborhood exploration algorithms as before except that each newly sampled genotype is added to the neutral network irrespective of system fitness. The neutral shadow is analyzed to show the neutral network properties for a maximally diffusive (i.e. unconstrained) neutral network. It provides an upper bound on both genotypic and topological distance measurements. Because the size and dimensionality of the degenerate and redundant fitness landscapes are identical, the neutral shadow generates the same topological and Hamming distance results for both system types.

## 5. Figures

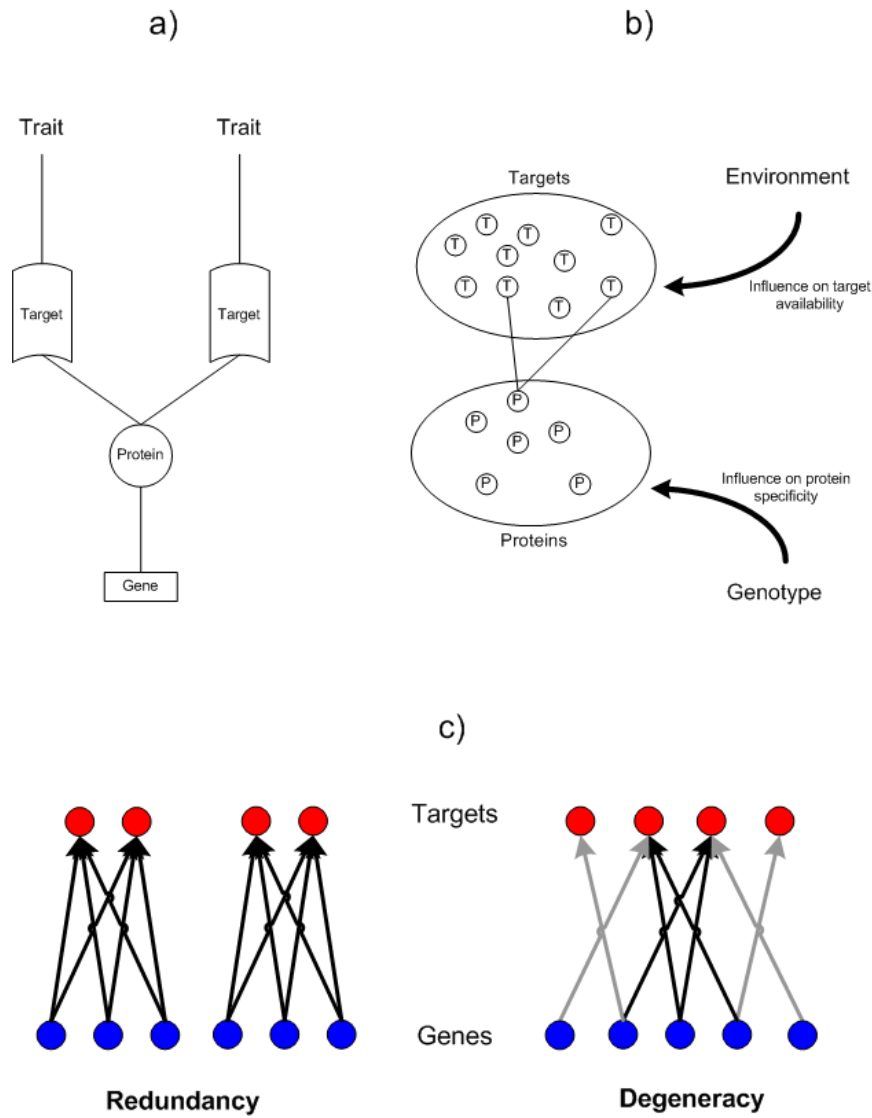

**Figure 1: Overview of genome-proteome model. a) Genotype-phenotype mapping conditions and pleiotropy:** Each gene contributes to system traits through the expression of a protein product that can bind with functionally relevant targets (based on genetically determined protein specificity). **b) Phenotypic expression:** Target availability is influenced by the environment and by competition with functionally redundant proteins. The attractor of the phenotype can be loosely described as the binding of each target with a protein. **c) Functional overlap of genes:** Redundant genes can affect the same traits in the same manner. Degenerate traits only have a partial similarity in what traits they affect.

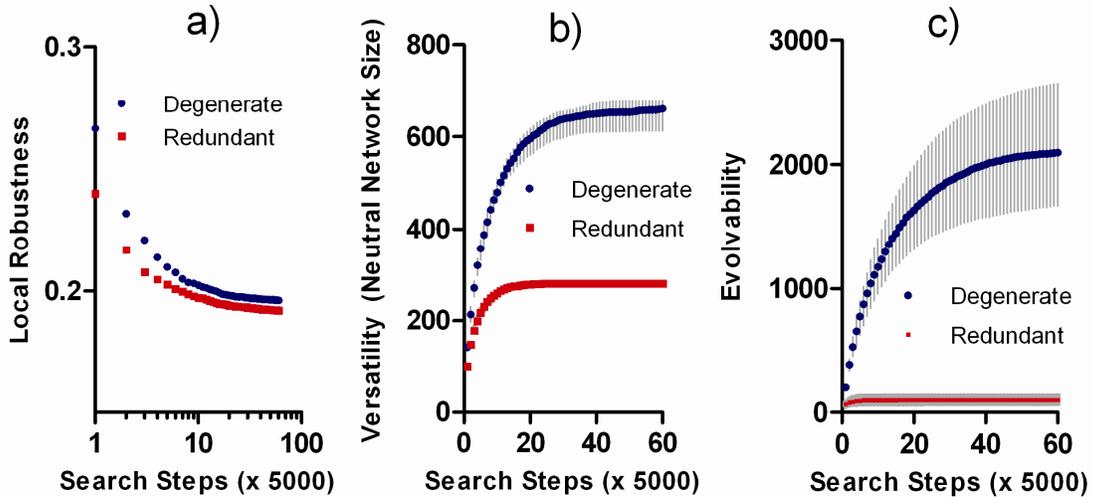

**Figure 2: Local robustness, versatility and evolvability measured as the fitness landscape is explored.** Each of the metrics is defined in the text and the procedure for exploring the fitness landscape is described in the Methods. Experiments are conducted with $m=8$ and $n=16$. Results show the median value from 50 runs with bars indicating 95% confidence intervals. For each position along the horizontal axis, degenerate and redundant data samples are found to be significantly different based on the Mann-Whitney U Test (U>2450, Umax = 2500, n1=50, n2=50) with larger median values in each case from the degenerate system ($p < 1E-6$).

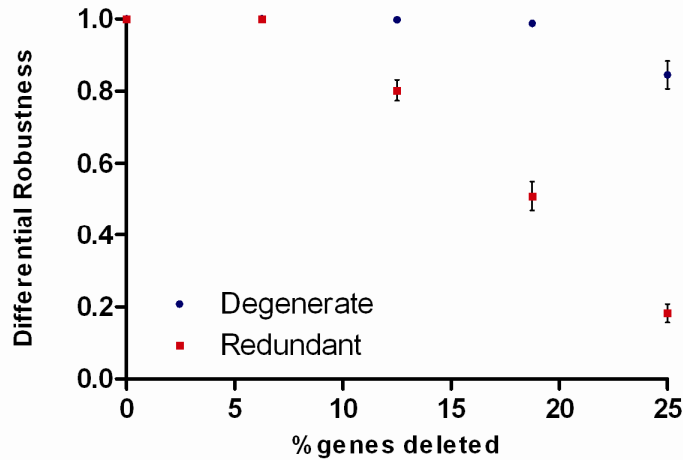

**Figure 3: Differential robustness of initial (un-mutated) systems as they are exposed to increasingly larger gene deletions.** Experiments are conducted with $m=8$ and $n=16$. Results are shown as the median value from 50 runs with bars indicating 95% confidence intervals.

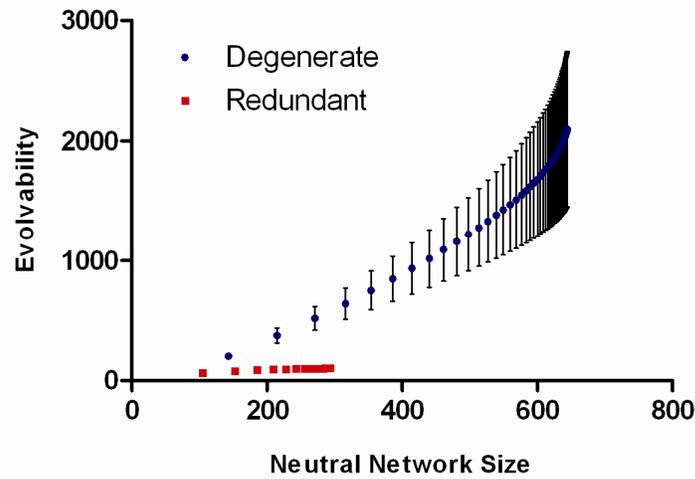

**Figure 4:** The number of unique phenotypes (evolvability) discovered versus the number of fitness neutral genotypes (neutral network) discovered. Similar behaviour is observed when evolvability is plotted against the size of the 1-neighborhood. Experiments are conducted with $m=8$ and $n=16$ and results are shown as the median value from 50 runs with bars indicating 95% confidence intervals.

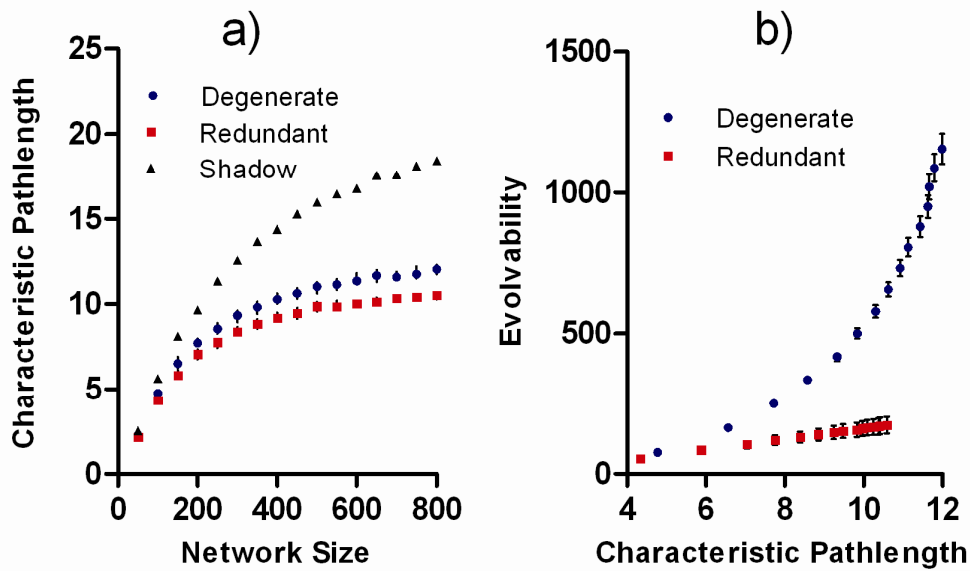

**Figure 5:** a) Characteristic path length of the neutral network for different network sizes ($n=8$, $m=20$, excess resources = 25%). The concept of resource excess is described in the context of Figure 7; it generates neutral networks larger than for those systems studied in Figure 4. Displayed results are medians of 50 experimental runs. Error bars indicate 95% confidence intervals; they are typically smaller than the resolution of data points. According to a Kruskal-Wallis test, characteristic path length distributions are significantly different ($p<1E-6$) for network sizes above 200. Results for the degenerate and redundant systems are also compared with a "shadow" of the neutral network search algorithm, which is described in Methods and provides an approximate upper bound for path length calculations. b) Evolvability as a function of characteristic path length.

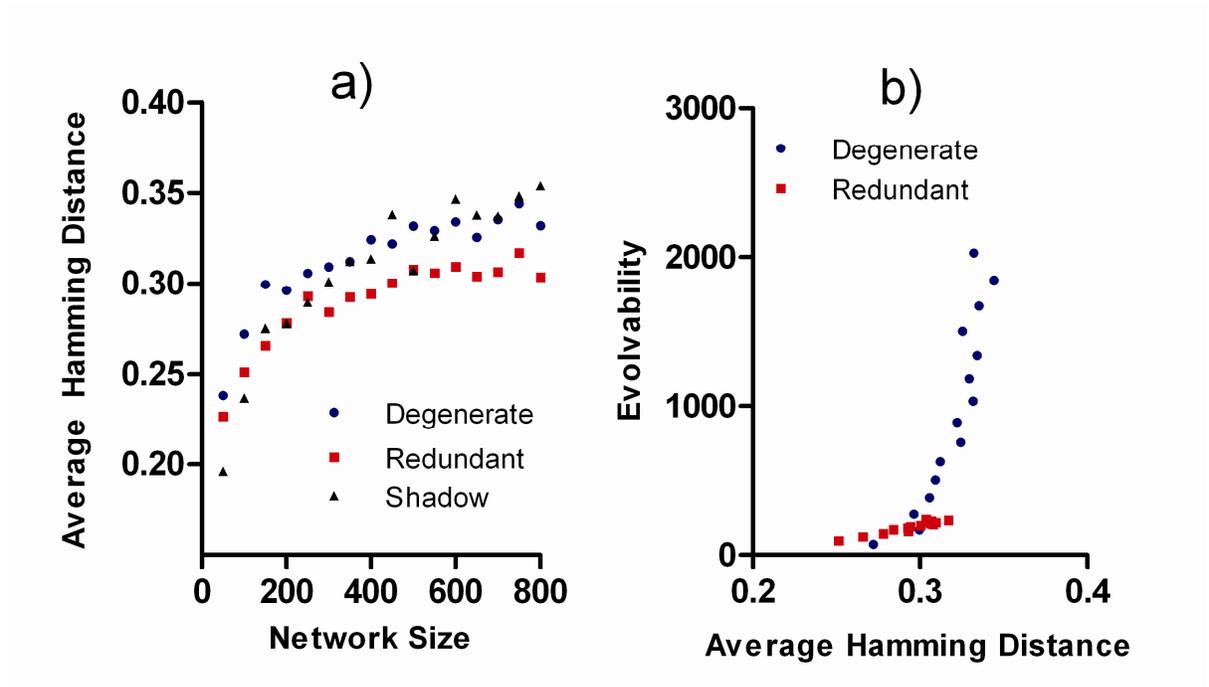

Figure 6: a) In genotype space, distance can be measured by the Hamming distance between genotypes. For the purposes of this measurement, genes are simply defined as binary values indicating whether each gene is deleted or not. Results are shown as the average Hamming distance of genotypes for all node pairs in the neutral network. Results are normalized with the maximum Hamming distance set equal to one. b) Evolvability is plotted as a function of the normalized Hamming distance. Similar results are obtained when analyzing the top 10% largest Hamming distances.

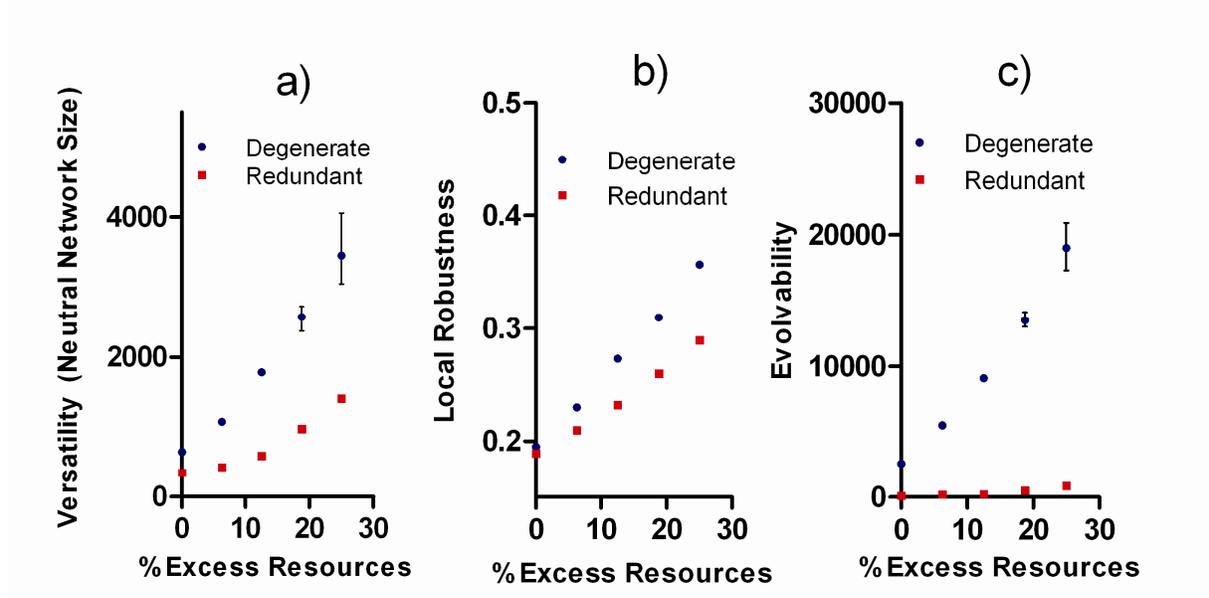

Figure 7: Versatility, robustness and evolvability as functions of excess components added to the two system types. For the baseline systems (i.e. systems without excess resources) the experiments are conducted with $m=8$, and $n=16$. For experiments where excess resources are larger than zero, the initial environment trait requirements $T^E$ are set based on the previous conditions ($m=8$, $n=16$) and then afterwards the system is redefined with $n$ increased by $n = 16*(1 + \%$ excess). Results represent the median value from 50 runs with bars indicating 95% confidence intervals. For each position along the horizontal axis, degenerate and redundant sample distributions are found to be significantly different based on the Mann-Whitney U Test (U>2450, Umax = 2500, n1=50, n2=50) with larger median values in each case from the degenerate system ($p < 1E-6$).

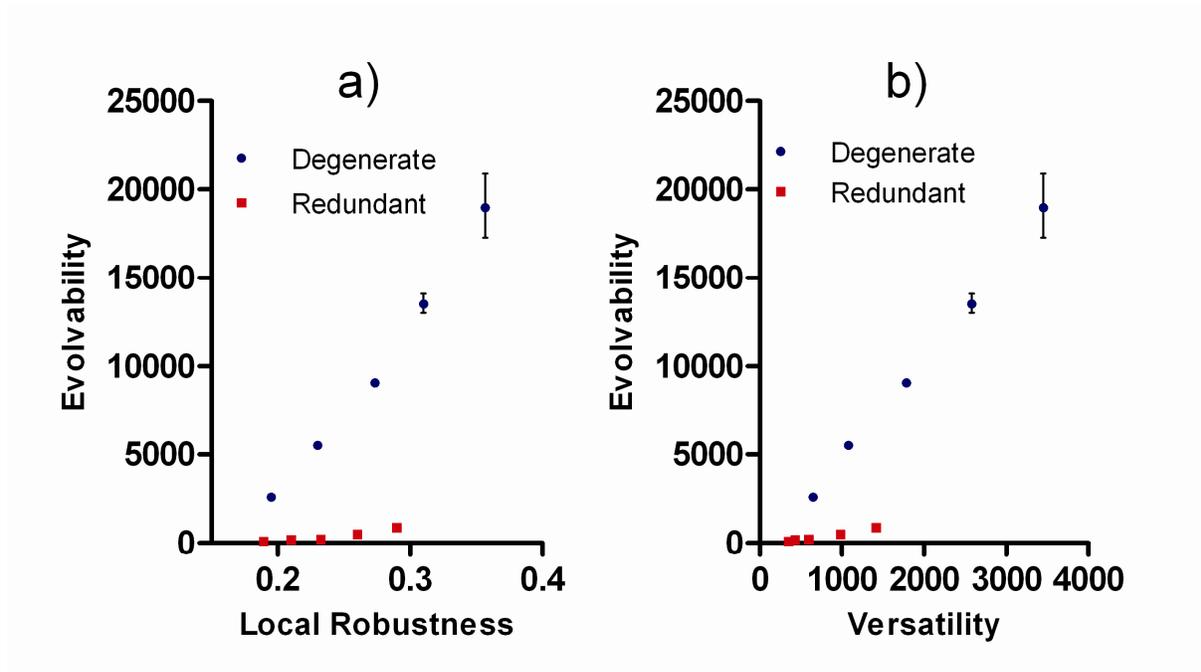

**Figure 8** Results of Figure 7 but with evolvability plotted as a function of local robustness (left) and versatility (right).

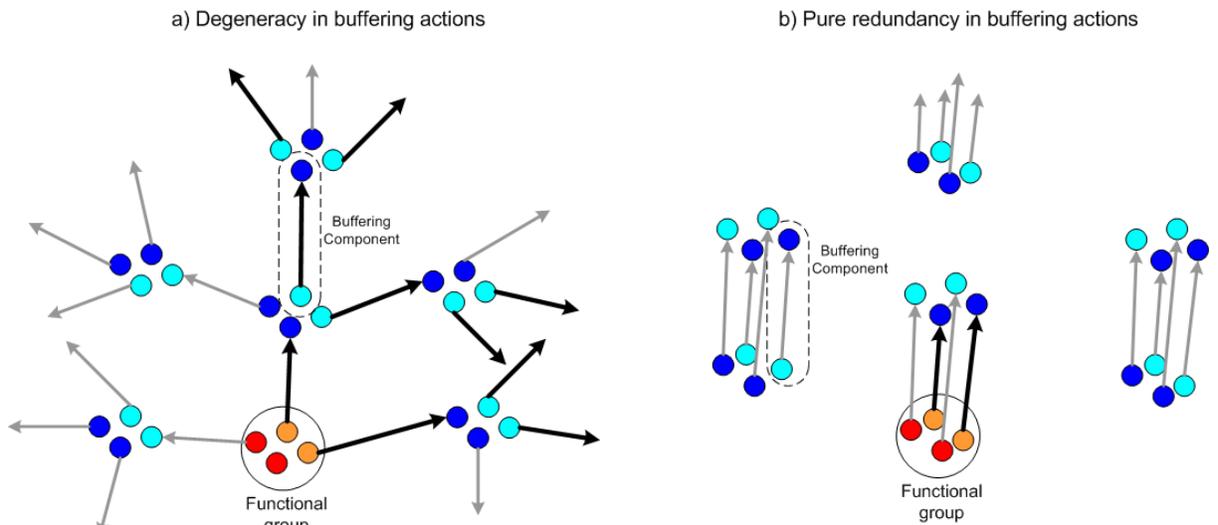

**Figure 9** Illustration of the connectivity of buffering actions provided by degeneracy. Functional groups are indicated by clusters of nodes, while connected node pairs represent individual components with a context dependent functional response (in this case, components have only two types of functional response). Dark/light shading is used to indicate which functional response a component is/is not currently carrying out. Darkened arrows indicate components that might be available if needed by the functional group from which the arrows originate. Here the darkened arrows illustrate how a stress to the circled functional group has the potential to cause a distributed response to that stress.

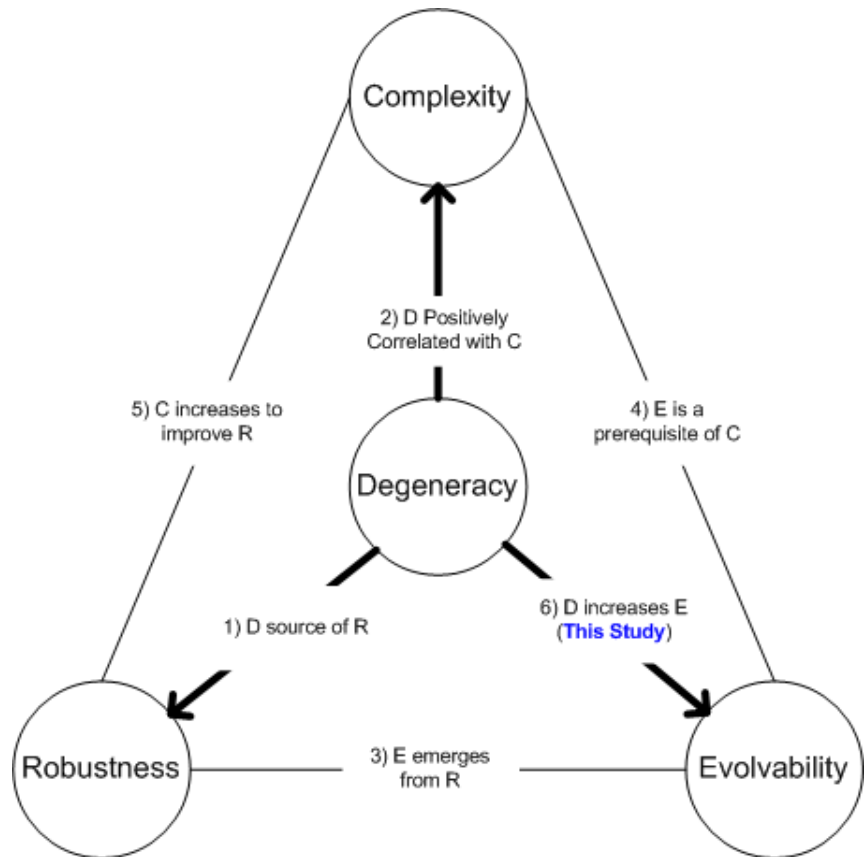

**Figure 10 Proposed relationship between degeneracy, evolution, robustness, and complexity**

# 6. Tables

**Table 1 Summary of evidence relating degeneracy, evolution, robustness, and complexity**

| | Relationship | Summary | Context | Ref |
|---|---|---|---|---|
| 1) | Degeneracy is a key source of biological robustness | Distributed robustness (and not pure redundancy) accounts for a large proportion of robustness in biological systems | Large scale gene deletion studies and other biological evidence (e.g. cryptic genetic variation) | [17] |
| 2) | Degeneracy has a strong positive correlation with system complexity | Degeneracy is correlated and conceptually similar to complexity. For instance degenerate components are both functionally redundant and functionally independent while complexity describes systems that are functionally integrated and functionally segregated. | Simulation models of artificial neural networks are evaluated based on information theoretic measures of redundancy, degeneracy, and complexity | [29] |
| 3) | Evolvability emerges from robustness | Genetic robustness reflects the presence of a neutral network. Over the long-term this neutral network provides access to a broad range of distinct phenotypes and helps ensure the long-term evolvability of a system. | Simulation models of gene regulatory networks and RNA secondary structure. | [8] [9] |
| 4) | Evolvability is a prerequisite for complexity | All complex life forms have evolved through a succession of incremental changes and are not irreducibly complex (according to Darwin's theory of natural selection). The capacity to generate heritable phenotypic variation (evolvability) is a precondition for the evolution of increasingly complex forms. | | |
| 5) | Complexity increases to improve robustness | According to the theory of highly optimized tolerance, complex adaptive systems are optimized for robustness to common observed variations in conditions. Moreover, robustness is improved through the addition of new components/processes that add to the complexity of the organizational form. | Based on theoretical arguments that have been applied to biological evolution and engineering design (e.g. aircraft, internet) | [31] [32] [33] |
| 6) | Degeneracy is a precondition for evolvability and a more effective source of robustness | Accessibility of distinct phenotypes requires robustness through degeneracy | Abstract simulation models of evolution | This Study |